\documentclass[draft,prd,aps]{revtex4}
\usepackage{amsmath,amssymb}
\usepackage{epsf}
\newcommand{\be}{\begin{equation}}
\newcommand{\ee}{\end{equation}}
\newcommand{\bea}{\begin{eqnarray}}
\newcommand{\eea}{\end{eqnarray}}
\begin{document}
\title{\bf Quantum Mechanics in Non(Anti)Commutative Superspace}
\author{L.G.~Aldrovandi and
F.A.~Schaposnik\footnote{F.A.S. is associated with CICBA.}}

\affiliation{Departamento\ de F\'{\i}sica, Universidad Nacional de
La Plata, C.C. 67, (1900) La Plata, Argentina.\\}

\begin{abstract}
We consider non(anti)commutative (NAC) deformations of $d=1$
${\cal N}=2$ superspace.  We find that, in the chiral base, the
deformation preserves only a half of the original (linearly
realized)  supercharge algebra, as it usually happens in NAC field
theories. We obtain in terms of a real supermultiplet a closed
expression for a deformed Quantum Mechanics Lagrangian in which
the original superpotential is smeared, similarly to what happens
for the two dimensional deformed sigma model.  Quite unexpectedly,
we find that  a second conserved charge can be constructed which
leads to a nonlinear field realization of the supersymmetry
algebra, so that finally the deformed theory has as many conserved
supercharges as the undeformed one. The quantum behavior of these
supercharges is analyzed.
\end{abstract}
\maketitle

\section{Introduction}

New aspects of supersymmetric field theories have been recently
analyzed by deforming superspace so that both bosonic and
fermionic coordinates obey non-trivial commutation and
anticommutation relations. Such deformed superspaces, first
proposed in \cite{Schwarz:1982pf}-\cite{deBoer:2003dn}, are known
as non(anti)commutative (NAC) superspaces and they have recently
attracted much attention both in connection with string theory
\cite{Billo:2004zq}-\cite{Aldrovandi:2006ea} and also with field
theories, like in studies of modified BPS solutions,  anomalies
and condensates  in SUSY gluodynamics, etc \cite{IS}-\cite{ASS}.

Supersymmetric Quantum Mechanics (SUSY
QM)\cite{Witten:1981nf}-\cite{Witten:1982df} provides a simple
and, at the same time, quite adequate framework for understanding
SUSY breaking non-perturbatively.  In particular, after the
introduction of the Witten index \cite{Witten:1982df}, many
 interesting  topological aspects
 of SUSY QM were clarified shedding light into more
 involved problem of SUSY
 breaking in quantum field theory.
For instance, the connection between the Witten index and QM
yields to a beautiful derivation
 of the
Atiyah-Singer index theorem
 \cite{Alvarez-Gaume:1983at}-\cite{Friedan:1983xr}, relevant for
analysis of   SUSY field theories.

In this work we study the NAC deformation of $d=1$ ${\cal N}=2$
superspace and the possibility of formulating Quantum Mechanics
 in such a superspace. Our purpose is twofold: on the one
hand we intend to fill a gap in the study of NAC theories by
analyzing the one dimensional case. We think that this  is
interesting by its own and for possible applications in  condensed
matter problems. On the other hand, as the study of QM is simpler
than that of  field theories, we shall use it  to analyze an
important issue of NAC theories, namely  the question of how many
supercharges are effectively broken by the deformation.   We know
that in NAC field theories, once the deformation is implemented,
only a fraction (usually a half) of the original linearly realized
algebra is left. Remarkably, we shall see that in the Quantum
Mechanics case a new, deformed supercharge, can be constructed (at
least classically), which corresponds to a non-linear realization
of the supersymmetry algebra.  This implies that the deformed
theory finally has as many conserved supercharges as the
undeformed one.

The paper is organized as follows. In section II we present our
definition of the NAC superspace and, working in the chiral base,
we show  that   the deformation preserves only half of the
supercharge algebra.  Then, in section III we  construct a
deformed QM Lagrangian by replacing the usual product between
superfields by the Moyal-Weyl product. Adapting  results obtained
in \cite{hatanaka1}-\cite{hatanaka2} for the $d=2, {\cal N}=2$ NAC
sigma-model, we find a $d=1$  Lagrangian which, written in
components, takes the same form as the undeformed one but with an
effective superpotential that depends not only on a scalar but
also on the auxiliary field. We then present two charges,
conserved in the deformed model, which lead to a nonlinear field
realization of the ${\cal N}=2$ superalgebra. In section IV we
study quantum aspects of the deformed theory. Using a
path-integral version of the Witten index ${\cal W}$ we show that
whenever the superpotential is such that SUSY is not broken in the
undeformed model, ${\cal W}$ remains unchanged by the deformation
and then the difference between the number of boson and fermion
states annihilated by the supercharges can be determined. Section
V is devoted to the formulation of the deformed theory using
chiral supermultiplets, this avoiding the use of auxiliary fields.
However, in order to preserve the chirality condition the class of
deformation turns to be restricted in a way such that the
resulting action coincides with the undeformed one. We present in
section VI a summary and discussion of our results.

\section{${\cal N}=2$ deformed superspace}

In the superspace formulation of supersymmetric Quantum Mechanics,
  apart from the (commuting) real time
variable $\tau$, one considers anticommuting (Grassmann) variables
$\theta$ and its complex conjugated $\bar \theta$. A deformed
superspace can be then defined by modifying the anticommutation
relations among the $\theta$ and $\bar \theta$ variables and even
their commutation relations with $\tau$.

In the ordinary case, the ${\cal N}=2$, $d=1$ superspace ${\mathbb
R}^{(1|2)}$ can be parameterized by the coordinates
\be
   {\mathbb
   R}^{(1|2)}=(\tau,\theta,\bar\theta),\;\;\tau\in\mathbb{R},\;(\theta)^\dagger=\bar\theta.
\ee
The most general deformation of the fermionic sector of this
superspace is
\be
    \{\theta,\theta\} = {C}, \;  \;\;\; \; \;\;\;
    \{\bar \theta,\bar \theta\} = \bar C,
    \;  \;\;\;  \;  \;\;\; \{\theta,\bar \theta\} = \widehat{C}
    \label{def1}
\ee
where ${C}$, $\bar C$ and $\widehat C$ are c-numbers having
dimensions ${\rm m}^{-1}$ since we take $[\theta] = [\bar\theta] =
{\rm m}^{-1/2}$. Concerning the commutation rules for the time
variable with $\theta$ and $\bar \theta$, we postpone the choice
to the introduction of a ``chiral''  time variable (see below).

As in the case of space-time deformations,  a $d=1$  theory in the
non(anti)commutative superspace defined by (\ref{def1}) can be
realized using ordinary Grassmann coordinates but multiplying
superfields with an appropriate Moyal-Weyl product. Indeed, given
for example scalar superfields of the form
\be
    \Phi (\tau,\theta,\bar\theta) = \phi(\tau) +  \theta\psi (\tau)+ \bar \psi(\tau)
    \bar \theta +  \theta \bar\theta F (\tau)
    \label{phi-1}
\ee
with $\phi$ a scalar, $\psi$ a fermion and $F$ an auxiliary field,
one can define the Moyal-Weyl product which implements the
deformation (\ref{def1}) as
\bea \Phi_1 (\tau,\theta,\bar\theta)*
     \Phi_2(\tau,\theta,\bar\theta)
    &= &\Phi_1 \exp(-\frac {{C}} {2}
    \frac{\overleftarrow{\partial}}{\partial \theta} \frac{
    \overrightarrow{\partial}}{\partial \theta} -
    \frac {{\widehat C}} {2}
    \frac{\overleftarrow{\partial}}{\partial \theta} \frac{
    \overrightarrow{\partial}}{\partial \bar\theta}-\frac {{\widehat C}} {2}
    \frac{\overleftarrow{\partial}}{\partial \bar\theta} \frac{
    \overrightarrow{\partial}}{\partial \theta}-\frac {\bar C} {2}
    \frac{\overleftarrow{\partial}}{\partial \bar\theta} \frac{
    \overrightarrow{\partial}}{\partial \bar\theta})
    \Phi_2\nonumber\\
    &= &\Phi_1
    \left[1-\frac {{C}} {2}
    \frac{\overleftarrow{\partial}}{\partial \theta} \frac{
    \overrightarrow{\partial}}{\partial \theta} -
    \frac {{\widehat C}} {2}
    \frac{\overleftarrow{\partial}}{\partial \theta} \frac{
    \overrightarrow{\partial}}{\partial \bar\theta}-\frac {{\widehat C}} {2}
    \frac{\overleftarrow{\partial}}{\partial \bar\theta} \frac{
    \overrightarrow{\partial}}{\partial \theta}-\frac {\bar C} {2}
    \frac{\overleftarrow{\partial}}{\partial \bar\theta} \frac{
    \overrightarrow{\partial}}{\partial \bar\theta}
    - \frac 1 4 \, c^2 \,\frac{\overleftarrow{\partial}}{\partial \bar\theta}
    \frac{\overleftarrow{\partial}}{\partial \theta}\frac{
    \overrightarrow{\partial}}{\partial \theta} \frac{
    \overrightarrow{\partial}}{\partial \bar\theta}\right]
    \Phi_2
    \label{bariloche}
\eea
with $c^2= {C}\bar C-{\widehat C}{\widehat C}$ and $\Phi
\frac{\overleftarrow{\partial}}{\partial \theta}=
(-1)^{f[\Phi]}\frac{\partial \Phi}{\partial \theta}$, where
$f[\Phi]$ is the Grassman character of the superfield $\Phi$.

In the undeformed  case, the differential operators $Q$ and $\bar
Q$ generating the supersymmetry transformations are given by
\be
   Q=\frac{\partial}{\partial\theta} +i\bar\theta\frac{\partial}{\partial \tau},\;\;\;\;\;\;\;\;\;\;\;\;
   \bar Q=-\frac{\partial}{\partial\bar\theta} -i\theta\frac{\partial}{\partial
   \tau},
   \label{cargas}
\ee
and satisfy the undeformed SUSY ${\cal N}=2$ QM algebra
\be
   \{Q,Q\}=0,\;\;\;\;\;\;\;\{\bar Q,\bar Q\}=0,\;\;\;\;\;\;\;
   \{Q,\bar Q\}=-2i\frac{\partial}{\partial \tau}.
\ee
It is sometimes useful to work with the supercharges realizations
\be
   Q=\frac{\partial}{\partial\theta} +ib\bar\theta\frac{\partial}{\partial t},\;\;\;\;\;\;\;\;\;\;\;\;
   \bar Q=-\frac{\partial}{\partial\bar\theta} -i\bar b\theta\frac{\partial}{\partial
   t}.
\ee
which can be obtained from (\ref{cargas}) through the change of
variables
\be \tau \rightarrow t=\tau- i(1-b)\theta\bar\theta \label{tiempo}
\ee
 and
satisfy the familiar supersymmetry algebra when $b+ \bar b=2$.
Concerning the covariant derivatives $D$ and $\bar D$
anticommuting with the supersymmetry charges, they take the form
\be
   D=\frac{\partial}{\partial\theta} -i\bar b\bar\theta\frac{\partial}{\partial t},\;\;\;\;\;\;\;\;\;\;\;\;
   \bar D=-\frac{\partial}{\partial\bar\theta} +i b\theta\frac{\partial}{\partial
   t}.
\ee

It is worthwhile to stress that we shall always consider the same
formal expression for the Moyal-Weyl product (\ref{bariloche}),
independently of having changed the temporal variable according to
(\ref{tiempo}). This means   that after the change $\tau
\rightarrow t$,    derivatives with respect to $\theta$ and
$\bar\theta$ in (\ref{bariloche}) should be taken at fixed $t$.

In the deformed case, the algebra of the supercharges $Q$ and
$\bar Q$ changes to
\bea
   &&\{Q,Q\}=-b^2 \bar C\frac{\partial^2}{\partial t^2}\nonumber\\
   &&\{\bar Q,\bar Q\}=-\bar b^2 {C}\frac{\partial^2}{\partial t^2}\nonumber\\
   &&\{Q,\bar Q \}=-i(b+ \bar b)\frac{\partial}{\partial t}+ b\bar b {\widehat C}
   \frac{\partial^2}{\partial t^2}
   \label{alge}
\eea
Different choices for $b$, $\bar b$ and the $C$-parameters make
possible to preserve different sectors of the undeformed SUSY
algebra. To select suitable values for these parameters, it is
convenient to see under which conditions the operators $Q$ and
$\bar Q$ still satisfy the Leibnitz rule. When $Q$ and $\bar Q$
act on $*$ product of superfields, one has
\bea
   Q*(\Phi_1*\Phi_2) &=& (Q*\Phi_1)*\Phi_2 + \Phi_1*(Q*\Phi_2)
   +ib[ \bar\theta \frac{\partial}{\partial t}(\Phi_1*\Phi_2)
   -(\bar \theta\frac{\partial\Phi_1}{\partial t})*\Phi_2
   -\Phi_1*(\bar \theta\frac{\partial\Phi_2}{\partial t})] \nonumber\\
   \bar Q*(\Phi_1*\Phi_2) &=&  (\bar Q*\Phi_1)*\Phi_2 +
   \Phi_1*(\bar Q*\Phi_2) -i\bar b[\theta \frac{\partial}{\partial t} (\Phi_1*\Phi_2)
   -(\theta\frac{\partial\Phi_1}{\partial t})*\Phi_2
   -\Phi_1*(\theta\frac{\partial\Phi_2}{\partial t})]
\eea
One can see that both supersymmetry generators cannot satisfy
simultaneously the Leibnitz rule  due to the constraint $b + \bar
b =2$. From here on we shall work in the chiral base, which
corresponds
 to the  $b=0,\, \bar b=2$ case and
then, according to (\ref{alge}), only the ${\cal N}=\frac 2 2$
SUSY subalgebra associated to the $Q$ generator will be preserved.
Let us also note that   this election is consistent with
\be [t,\theta] = [t,\bar \theta] = 0 \ee
which corresponds, in terms of the   original coordinates
$(\tau,\theta,\bar\theta)$,   to
\be
   [\tau, \tau]=0,\;\;\;\;\;\;\;[\tau,\theta]=i(\hat C \theta - C\bar\theta)
   ,\;\;\;\;\;\;\;[\tau,\bar\theta]=i(\bar C \theta - \hat
   C\bar\theta)   \, .
   \label{falk}
\ee
Finally, the  supercharge and covariant derivatives take the form
\bea
  Q&=&\frac{\partial}{\partial\theta} \; ,  \;\;\;\;\;\; \;\;\;\;\;\;\;\;\;\;\;
   \bar Q=-\frac{\partial}{\partial\bar\theta}
   - 2i\theta\frac{\partial}{\partial t},\nonumber\\
     D&=&\frac{\partial}{\partial\theta} - 2i\bar\theta\frac{\partial}{\partial t}\; ,
     \;\;\;\;\;\;\;
   \bar D=-\frac{\partial}{\partial\bar\theta}   \, .
   \label{charges}
\eea

\section{Lagrangian and classical supersymmetries}

There are two basic kinds of ${\cal N} = 2$ supermultiplets in
$d=1$ dimension. To classify them it is convenient to introduce
the notation ${\bf (m, n, n -m)}$ to identify an off-shell ${\cal
N} = {\bf n}$, $d = 1$ supermultiplet with ${\bf m}$ physical
bosons, ${\bf n}$ fermions and ${\bf n -m}$ auxiliary bosonic
components. Thus, the ${\bf (1,2,1)}$ supermultiplet, also known
as ${\cal N} = 2a$ model, is constructed from unconstrained real
${\cal N} = 2$ superfields while the ${\bf (2,2,0)}$
supermultiplet or ${\cal N} = 2b$ model use complex chiral
superfields. In the undeformed superspace, ${\cal N} = 2a$ models
can be obtained by dimensional reduction of $(1,1)$ supersymmetric
two dimensional sigma-models, while the reduction of $(2,0)$
supersymmetric two dimensional sigma-models leads to ${\cal N} =
2b$ models.

A deformed ${\cal N}=2$ supersymmetric action constructed by using
the ${\bf (1,2,1)}$ supermultiplet takes the form
\bea
    S &=& \int dt d \theta d\bar \theta \left[ -\frac{1}{2} (D*\Phi)*
    (\bar D* \Phi) - W_*(\Phi) ] \right)
\eea
where one defines the deformed superpotential $W_*(\Phi)$ from the
function $W(\Phi)$ in the underformed case through
\be
   W(\Phi)=\sum_{n=0}^\infty \frac {1}{n!}W^{(n)}\,\Phi
  \Phi \stackrel{(n)}{\ldots}\Phi\longrightarrow
   W_*(\Phi)=\sum_{n=0}^\infty \frac {1}{n!}W^{(n)}\,
   \Phi*\Phi*\stackrel{(n)}{\ldots}*\Phi
\ee
where the superfield $\Phi$ has the component expansion
(\ref{phi-1}),
\be
    \Phi (t,\theta,\bar\theta) = \phi(t) +  \theta\psi (t)+ \bar \psi(t)
    \bar \theta +  \theta \bar\theta F (t)
\ee
In terms of the $(\tau, \theta,\bar \theta)$ coordinates, the
superfield $\Phi$ is real, that is,
$\phi^\dagger(\tau)=\phi(\tau)$,
$\psi^\dagger(\tau)=\bar\psi(\tau)$ and $F^\dagger(\tau)=F(\tau)$.
However, the change of variables $\tau\rightarrow t$ complexifies
the temporal coordinate and then, the  reality relations for the
component fields are lost.

In the two-dimensional NAC ${\cal N}=2$ sigma model context
 Alvarez-Gaum\'e and V\'azquez-Mozo found
a remarkable non-perturbative formula for the deformed holomorphic
superpotential \cite{hatanaka1}-\cite{hatanaka2}. Now, this
formula can be also applied in this $d=1$ case and allows us to
express the higher component of the superpotential as
\be
   \int d\theta d\bar\theta\, W_*(\Phi) =  F \frac{\partial}{\partial \phi}
   \int_{-\frac 1 2}^{+\frac 1 2}d\xi \,W(\phi+\xi c F)- \bar\psi \psi\frac{\partial^2}{\partial \phi^2}
   \int_{-\frac 1 2}^{+\frac 1 2}d\xi\, W(\phi+\xi c F)
   \label{pote}
\ee
with
\be
   c=\sqrt{{C}\bar C-{\widehat C}{\widehat C}}.
   \label{cee}
\ee
As noted in \cite{hatanaka1}-\cite{hatanaka2}, this deformation
corresponds, physically, to a smearing of the target space
coordinates. According to this, an effective superpotential
$\tilde W$ can be obtained by averaging its undeformed value
between $\phi-c F$ and $\phi+ c F$, that is \be
   \tilde W(\phi, F) =  \int_{-\frac 1 2}^{+\frac 1 2}d\xi \,W(\phi+\xi c F).
\ee
So, written in components,   the deformed action for Quantum
Mechanics
 takes the form
\be
   S = \int dt\, \left [-\left(i\frac{d\phi}{dt} + \frac{\partial \tilde W}{\partial \phi} \right)
   F + \frac 1 2 F^2 +  \bar\psi\left(i\frac{d}{dt} + \frac{\partial^2 \tilde W}{\partial \phi^2} \right)
   \psi\right]
   \label{action1}
\ee

In order to study classical aspects of this deformed theory let us
use the usual definition for the momenta, Hamiltonian and Poisson
brackets
\bea
    &&\Pi^i = \frac{\partial_r{\cal L}}{\partial \dot
    \Theta_i},\;\;\;\;\;\;\;\; {\cal H}=\Pi^i\dot
    \Theta_i-{\cal L},\;\;\;\;\;\;\;\; \{X,Y\}_P=\frac{\partial_r X}{\partial \Theta_i}\frac{\partial_l Y}{\partial
    \Pi^i}- (-1)^{XY}\frac{\partial_r Y}{\partial \Theta_i}\frac{\partial_l
    X}{\partial \Pi^i},
\eea
where the subscripts $r$ and $l$ denotes right and left
derivatives, respectively. Hence, with $\Theta=(\phi,\psi)$, the
conjugate momenta are $\Pi=(-iF,i\bar\psi)$ and the Hamiltonian
associated with the deformed action (\ref{action1}) is
\bea
   {\cal H} = -iF\dot \phi + i \bar\psi\dot\psi - {\cal L}=
   \frac{\partial \tilde W}{\partial \phi}F - \frac 1 2 F^2 -
    \bar\psi\frac{\partial^2 \tilde W}{\partial \phi^2} \psi
   \label{jamil}
\eea
One can easily  check that the action (\ref{action1}) is still
invariant under the supersymmetry generated by $Q$
\be
   \delta_Q\phi=\psi,\;\;\;\;\;\;\;\delta_Q\psi=0,\;\;\;\;\;\;\;
   \delta_Q\bar\psi=F,\;\;\;\;\;\;\;\delta_Q F=0,
   \label{vari}
\ee
where the variations for the fields are obtained from
\be
   \delta_Q \Phi  =Q*\Phi= \delta_Q\phi-  \theta\delta_Q\psi +  \delta_Q\bar\psi
    \,\bar \theta +  \theta \bar\theta\delta_Q F.
\ee
On the other hand, we have seen that the differential operator
$\bar Q$ as defined in (\ref{charges}) and satisfying  $\{\bar Q,
\bar Q\}\neq 0$, cannot be used as a generator of SUSY
transformations. Besides,  the $\bar Q$-variation of the component
fields corresponding to the undeformed case are no more   a
symmetry of the deformed theory. That is, given the variations
\be
   \delta_{{\bar Q}_{c=0}}\phi=\bar\psi,\;\;\;\;\;\;\;\delta_{{\bar Q}_{c=0}}\psi=2i\dot \phi - F,\;\;\;\;\;\;\;
   \delta_{{\bar Q}_{c=0}}\bar\psi=0,\;\;\;\;\;\;\;\delta_{{\bar Q}_{c=0}} F=2i\dot{\bar\psi},
   \label{variraya}
\ee
the Lagrangian does not remain invariant but changes according to
\be
   \delta_{{\bar Q}_{c=0}}{\cal L}= 2i \bar\psi\frac{\partial^2 \tilde W}{\partial F\partial
   \phi}\dot F - 2i \dot{\bar\psi}\frac{\partial^2 \tilde W}{\partial F\partial
   \phi} F - 2i \bar\psi\dot{\bar\psi}\frac{\partial^3 \tilde W}{\partial F\partial
   \phi^2}\psi,
\ee
Then,  only for $c=0$ (the undeformed case)
  $\delta_{{\bar Q}_{c=0}}{\cal L}$   vanishes. This result was to be expected
from the form of the deformed algebra of the supercharge
(\ref{alge}) which, after taking $b=0,\,\bar b=2$, is broken in
the $\bar Q$ sector. This  is similar to what happens in NAC field
theories where at least half of the original supersymmetry is
broken due to the deformation.

Coming back to the supersymmetry that is preserved,  we can give
an expression of the supercharge $Q$ in terms of the fields
following the Noether prescription,
\be
    {\cal Q}=-\sum_\chi \frac{\partial_r{\cal L}}{\partial \dot
    \chi}\delta_Q\chi = i\psi F,
\ee
(with $\chi$ we denote the component fields in the theory). One
can see that ${\cal Q}$  satisfies \be \{{\cal Q}, {\cal Q}\}_P=0
\, , \;\;\; \;\;\; \{{\cal Q}, \chi\}_P=\delta_Q \chi \label{refe}
\ee
for the variations (\ref{vari}). One can then check that the
Hamiltonian is ${\cal Q}$-exact since
\be
   {\cal H} = \left\{{\cal Q}, \frac 1 2 \bar\psi\left(-F + 2
   \frac{\partial \tilde W}{\partial \phi}\right)\right\}_P
\ee
This expression suggests the existence of a second charge
$\bar{\cal Q}$ so that, as in the undeformed case,  the deformed
Hamiltonian could be written  as
\be
   {\cal H}=\frac i 2 \{{\cal Q}, \bar{\cal Q}\}_P.
   \label{eli}
\ee
This can be achieved by defining the charge $\bar{\cal Q}$ so
that, in terms of component fields, it takes the form
\be
    \bar{\cal Q}= i\bar\psi\left(F -2
   \frac{\partial \tilde W}{\partial \phi}\right)
\ee
Surprisingly,  the charge $\bar{\cal Q}$ is conserved,
\be \{{\cal H},\bar{\cal Q}\}_P=0 \ee and it is also nilpotent,
\be \{\bar{\cal Q}, \bar{\cal Q}\}_P=0 \ee

We conclude that, at least classically, the deformed theory has as
many supercharges as the undeformed one. Acting on the fields,
 the second charge  gives the transformations
\bea
    &&\{\bar{\cal Q}, \phi\}_P= \bar\psi
    - 2\bar\psi\frac{\partial^2 \tilde W}{\partial \phi\partial
    F},\;\;\;\;\;\;\;\;\;\;\;\;\;\;\{\bar{\cal Q}, \bar\psi\}_P=0,\nonumber\\
    &&\{\bar{\cal Q}, \psi\}_P= F - 2\frac{\partial \tilde W}{\partial \phi},
    \;\;\;\;\;\;\;\;\;\;\;\;\;\;\;\;\;\;\;
    \{\bar{\cal Q}, F\}_P= 2\bar\psi\,\frac{\partial^2 \tilde W}{\partial
    \phi^2}.
    \label{2}
\eea
Note that in the $c \to 0$ limit $\bar{\cal Q} \to {\bar
Q}_{c=0}$, as one can verify by comparing transformations
(\ref{variraya}) and (\ref{2}) and using component fields
equations of motion. It is important to stress that, in contrast
with the case of the transformations (\ref{variraya}), which gives
a linear realization of the supersymmetry algebra in the
undeformed theory, $\bar{\cal Q}$ provides, together with ${\cal
Q}$, a non-linear realization defined by
eqs.(\ref{refe}),(\ref{2}).

\section{Quantum supersymmetries}

To study the quantum theory, one should replace the Poisson
bracket by commutators, $\{\; ,\,\}_P\rightarrow -i[\,,\,]$ and
write the canonical commutation relations
\be
   [\phi, \Pi_\phi]=i,\;\;\;\;\;\;\;\{\psi,i\bar\psi\}=1.
\ee
to describe the Hilbert space. The first relation can be
represented in the standard way in terms of wavefunctions in
$L^2(\mathbb{R})$. To represent the second one, which corresponds
to a Clifford algebra, we can use the $2\times 2$ matrices, \be
   \psi=\left(
   \begin{array}{cc}
   0 & 1\\
   0 & 0
   \end{array}\right),\;\;\;\;\;\;\;\;\;\;\;\;\;\;
   i\bar\psi=\left(
   \begin{array}{cc}
   0 & 0\\
   1 & 0
   \end{array}\right).
\ee

In order to study  symmetries at the quantum level it will be more
convenient to work with a new charge
\be \hat{\cal Q}_+=\hat{\cal Q}+\hat{\bar{\cal Q}} \ee (or
equivalently with $\hat{\cal Q}_-=i(\hat{\cal Q}-\hat{\bar{\cal
Q}})$) which satisfies $\hat{\cal H}=(\hat{\cal Q}_+)^2$. Note
that this relation imposes a constraint to the possible orderings
of $\hat{\cal H}$ and $\hat{\cal Q}_+$.

In order to determine whether the ${\cal Q}$-supersymmetries are
spontaneously broken in the deformed theory, one can analyze the
Witten index. Since the Hamiltonian is the square of $\hat{\cal
Q}_+$, then  $[\hat{\cal H},\hat{\cal Q}_+]= 0$. Thus, except for
states annihilated by $\hat{\cal Q}_+$, a  boson/fermion pairing
of states with the same energy takes place and the Witten index
results
\bea
    {\cal W} ={\rm Tr} \left[(-1)^F e^{-\beta {\cal H}}\right] = \{\sharp\;
    {\rm bosonic}\; \Upsilon_+^{\rm closed\; states}\} -
    \{\sharp\; {\rm  fermionic}\; \Upsilon_+^{\rm closed\; states}\}
\eea
where $\Upsilon_+^{\rm closed\; states}$ are those states which
are annihilated by $\hat{\cal Q}_+$.

Were $\hat{\cal Q}_+$ an Hermitian operator, the following
identity would be valid
\be
   \langle\chi\mid\hat{\cal H}\mid\chi\rangle=\mid\hat{\cal
   Q}_+\mid\chi\rangle\mid^2,
\ee
and then states annihilated by $\hat{\cal Q}_+$ would be vacuum
states. As a consequence, a non vanishing Witten index would imply
that the ${\cal Q}_+$-symmetry is not broken. This is in fact what
happens for supersymmetric theories in  undeformed superspace.
Now, when the deformation is present the situation radically
changes: as it happens for non(anti)commutative field theories,
the action, and hence the Hamiltonian, are in general complex.
Then, according to (\ref{eli}), $\hat {\bar{\cal Q}}$ is not the
hermitian
 conjugate of $\hat {{\cal Q}}$, this leading in general to non-hermitian
 charges $\hat{\cal Q}_{\pm}$.

Now,  it has been proved \cite{chec} that the Witten index can be
obtained from the partition function $Z$ in Euclidean time through
standard manipulations,
\be
   {\cal W}=\int_{PBC} {\cal D}\Theta\,{\cal D}\Pi\, \exp\left (-{\int_0^\beta dt\, (\Pi^i\dot
    \Theta_i-{\cal H})}\right),
    \label{w}
\ee
with $\Theta=(\phi,\psi)$, $\Pi=(iF,-i\bar\psi)$. Canonical
variables in the Hamiltonian operator should be orderer so that
they satisfy ${\cal H}=e^{ip_\phi\phi}\langle p_\phi|\hat{\cal
H}|\phi\rangle$ and the functional integral is taken over fields
satisfying periodic boundary conditions:
$\Theta(t+\beta)=\Theta(t)$. With this representation for the
Witten index, its calculation can be easily performed since the
deformed partition function coincides with the undeformed one when
the ${\cal Q}$-symmetry is not broken in the undeformed $(c=0)$
theory. To see this we should first note that the deformed action
is ${\cal Q}$-exact, that is,
\be
   S=\int dt\, \left\{{\cal Q},\bar\psi \left(\frac{d\phi}{dt} +
   \frac{\partial \tilde W}{\partial \phi} - \frac i 2
   F\right)\right\}_P = S(c=0) + \int dt\, \left\{{\cal Q},\bar\psi
    \frac{\partial }{\partial \phi}(\tilde W-W)\right\}_P
\ee
As a result, we can express the partition function as
\bea
    Z&=&\int {\cal D}\Phi \exp(-S)= \int {\cal D}\Phi
    \exp\left(-S(c=0) - \int dt\, \left\{{\cal Q},\bar\psi
   \frac{\partial }{\partial \phi}(\tilde W -W) \right\}_P(t)\right)\nonumber\\
   &=& Z(c=0) + \sum_{n>0}(-1)^n \int dt_1\ ... \int dt_n\times\nonumber\\
   && \;\;\;\;\;\;\;\;\;\;\;\;\;\;\;\;\;\;
   \left\langle \left\{{\cal Q}_{c=0},\bar\psi
   \frac{\partial }{\partial \phi}(\tilde W -W) \right\}(t_1)...
   \left\{{\cal Q}_{c=0},\bar\psi\frac{\partial }{\partial \phi}(\tilde W -W)
   \right\}(t_n)\right\rangle_{c=0}
\eea
where $\langle\;\rangle_{c=0}$ denote the expectation value
corresponding to the undeformed case. Note that the replacement
 $\{{\cal Q},\;\}\rightarrow\{{\cal Q}_{c=0},\;\}$ can
be done since the ${\cal Q}$-transformations of the field
components are not affected by the deformation. If ${\cal
Q}_{c=0}$-symmetry is not broken in the undeformed case,
expectation values of ${\cal Q}_{c=0}$-exact operators vanish.
Thus we get that $Z=Z(c=0)$ and, due to relation (\ref{w}), ${\cal
W}={\cal W}(c=0)$ when the undeformed theory is supersymmetric.

We can summarize the latter results as follows,
\bea
 {\cal W} =   \{\sharp\;
    {\rm bosonic}\; \Upsilon_+^{{\rm closed\; states}}\} - \{\sharp\; {\rm  fermionic}\;
    \Upsilon_+^{\rm closed\; states}\} = \left\{
   \begin{array}{l}
   -1 \;\; \;\; {\rm for}\, W(\phi)\rightarrow a\phi^{2n}\,{\rm ~as}\, \phi
   \rightarrow \pm\infty,\;a>0 \\
   \hphantom{-}1 \;\; \;\;{\rm for}\, W(\phi)\rightarrow -a\phi^{2n}\,{\rm ~as}\,
   \phi\rightarrow \pm\infty \\
 \hphantom{-}  {\rm unknown\,in\,other\,case.}
   \end{array}\right.
\eea
Taking vacuum states as those for which the v.e.v. of $\hat{\cal
H}$ vanishes, the states annihilated by ${\cal Q}_+$ are vacuum
states. Then, those cases in which ${\cal W} \ne 0$ correspond to
theories with unbroken supersymmetry.

\section{The ${\bf (2,2,0)}$ supermultiplet}

When one formulates the deformed theory in terms of the ${\bf
(2,2,0)}$ supermultiplet, that is, by using complex chiral
superfields instead of real ones,  auxiliary fields are not
introduced. It could then be promising to analyze this case since
it was the presence of the auxiliary field which prevented a
complete answer to the SUSY symmetry breaking in the case of the
${\bf (1,2,1)}$ multiplet.

In the chiral base, a chiral superfield $\Psi$ and an antichiral
superfield $\bar\Psi$ have the component expansions
\bea
    \Psi(t,\theta)&=&z(t) + \theta\chi(t)\nonumber\\
    \bar\Psi(t-2i\theta\bar\theta,\bar\theta)&=&\bar
    z(t-2i\theta\bar\theta)
    + \bar\chi(t)\bar\theta=\bar z(t)
    + \bar\chi(t)\bar\theta-2i\theta\bar\theta\dot{\bar z}(t)
\eea
In terms of $(\tau,\theta,\bar\theta)$ coordinates, the component
fields satisfy $z^\dagger(\tau)=\bar z(\tau)$ and
$\chi^\dagger(\tau)=\bar \chi(\tau)$. Now, in order to use these
supermultiplets one needs the Moyal-Weyl product to preserve the
superfields chirality condition. With the generic deformation
(\ref{def1}), the products of chiral and antichiral superfields
take the form
\bea
    \Psi_1*\Psi_2&=&\Psi_1\Psi_2 -\frac
    {{C}}{2}\chi_1\chi_2\nonumber\\
    \bar\Psi_1*\bar\Psi_2&=&\bar\Psi_1\bar\Psi_2-\frac
    {\bar C}{2}\bar\chi_1\bar\chi_2 + c^2
    {\dot{\bar z}}_1{\dot{\bar z}}_2 + i\bar C
    \theta \left({\dot{\bar z}}_1\bar\chi_2-{\dot{\bar z}}_2\bar\chi_1\right)
    + i{\widehat C} \left({\dot{\bar z}}_1\bar\chi_2-{\dot{\bar z}}_2\bar\chi_1
    \right)\bar\theta
\eea
One can see that a product of chiral superfield is still chiral.
However, in order to preserve the antichirality condition, the
deformation has to satisfy $\bar C=c^2=0$. Then, only the $\theta$
algebra gets deformed.

Let us consider a theory with the kinetic energy canonically
normalized and a real K\"{a}hler prepotential $K(z,\bar z)$. The
deformed action is
\bea
    S &=& \int dt d \theta d\bar \theta \left[ \frac{1}{4} (D*\Psi)*
    (\bar D* \bar\Psi) + K_*(\Psi,\bar\Psi) ] \right),
    \label{action2}
\eea
with the deformed prepotential defined as
\be
   K_*(\Psi,\bar\Psi)=\sum_{n,m=0}^\infty \frac{1}{(n+m)!}
   \frac{\partial^{n+m}K}{\partial z^{n}
   \partial \bar z^{m}}\,[\Psi*\stackrel{(n)}{\ldots}*\Psi*
   \bar\Psi*\stackrel{(m)}{\ldots}*\bar\Psi]
\ee where the square brackets $[...]$ means all possible
permutations of the superfields.

A non-perturbative expression for the deformed K\"{a}hler
prepotential can be obtained using a generalization of
ec.(\ref{pote}) for several superfields
\cite{hatanaka1}-\cite{hatanaka2}. However, as in the case of one
superfield, the deformation of the prepotential is controlled by
the parameter $c^2$, which in this case is zero, and then the
prepotential does not receive corrections. As the kinetic term
neither is affected by the deformation, the action (\ref{action2})
results undeformed.

\section{Summary and discussion}

We have analyzed in this work  NAC deformations of the $d=1$
${\cal N}=2$ superspace and studied how a Quantum Mechanics model
can be defined on it. In particular, the emphasis was put on the
determination of how many supercharges were broken due to the NAC
deformation. Starting from a general deformation of the superspace
fermionic coordinates  and working in the chiral base, we found
that one half of the original supersymmetry algebra (corresponding
to the supercharge $Q$) can be preserved, the same as it happens
in NAC field theories.

By using a Moyal-Weyl product to realize the deformation, we
defined a deformed ${\cal N}=2$ SUSY theory in terms of the ${\bf
(1,2,1)}$ supermultiplet for an arbitrary superpotential. To do
this we applied a remarkable result
\cite{hatanaka1}-\cite{hatanaka2} for the deformed two dimensional
sigma model
 which allows  to obtain a closed expression for the
deformed superpotential. In this way we found that the resulting
Lagrangian takes the same form as the undeformed one but with a
superpotential which is the average of  its undeformed value in
the interval  $( \phi- cF/2,\phi + cF/2)$, with $c$ related to the
deformation as given by formula (\ref{cee}). As a consequence, the
NAC deformed superpotential depends not only on $\phi$ but also on
the auxiliary field $F$.

Concerning the supersymmetries of the deformed model, an
expression in terms of the component fields for the surviving
charge ${\cal Q}$ was easily found by using its differential
operator representation (\ref{charges}). The fact that the
Hamiltonian was still ${\cal Q}$-exact made natural   to propose
an ansatz for a second supercharge $\bar {\cal Q}$. This charge is
found to be conserved and   together   with ${\cal Q}$,   defines
a nonlinear realization of
 the supersymetry
algebra:
\be
   \{{\cal Q},\bar{\cal Q}\}_P=-2i{\cal H},\;\;\;\;\;\;\;\;\;
   \{{\cal Q},{\cal Q}\}_P=0,\;\;\;\;\;\;\;\;\;
   \{\bar{\cal Q},\bar{\cal Q}\}_P=0.
\ee
As a result, we have the unexpected result that the deformed model
has as many (conserved) supercharges as the undeformed one.

We analyzed the conservation of these supercharges in the quantum
theory and whether they were or not spontaneously broken. The fact
that the deformed Hamiltonian is, in general, non hermitian leads
to non-hermiticity of supercharges. However, we found that,
whenever the undeformed theory is supersymetric, the Witten index,
defined as a path-integral, does not change with the deformation.
This result allowed to determine the difference between the number
of boson and fermion states annihilated by the supercharge and
then, to know when the supercharge is not spontaneously broken.

Finally, we also formulated the deformed theory in terms of the
${\bf (2,2,0)}$ supermultiplet, that is, by using complex chiral
superfields instead of real ones. In this case, requiring  the
Moyal-Weyl product to preserve the chirality condition imposes
constraints on the class of deformations. Now these constraints
turn out to be too restrictive so that finally the resulting
deformed action coincides with the undeformed one.

Our work suggests that NAC quantum field theories could have more
supersymmetries than those one naively expect in view of the
deformation. It would be very interesting to gain a deeper insight
in this direction.

Non(anti)commutative Quantum Mechanics is also a very interesting
topic for further analysis in connection with the no-linear sigma
model.
 It is known that the addition of   more ``superspace structure'' by
means of NAC deformations leads to new deformations in complex
geometry, whose geometrical significance is yet to be understood
\cite{Ketov:2006zx}. The study of NAC deformed one dimensional
no-linear sigma model can shed some light on this issue.

Finally, as it happens with Noncommutative QM, non(anti)
commutative QM could have applications in condensed matter
problems. For instance, it has been shown that, in supermatrix
models, fuzzy superspheres arise as classical solutions, and their
fluctuations yield to NAC field theories \cite{Iso:2003zb}. Some
interesting relations between lowest Landau level (LLL) physics
and NAC geometry have also been reported
\cite{Hatsuda:2003ry},\cite{Ivanov:2003qq}. With these recent
developments, the supersymmetric quantum Hall systems might be the
simplest ``physical" set-up of NAC geometry.

\vspace{1.5 cm}

\noindent\underline{Acknowledgements}: F.A.S. wishes to thank L.
Cugliandolo, E. Fradkin, E. Moreno and J. Kurchan for their
comments and patience the many times they had to hear about the
ideas behind this work. L.A. is supported by CONICET. This work is
partially supported by UNLP, CICBA, CONICET (PIP 6160) and ANPCYT.

\end{document}